\documentclass[journal]{IEEEtran}
\ifCLASSINFOpdf
\else
\fi
\usepackage{float}
\usepackage{amsfonts,amsopn,dsfont} 
\usepackage{cite}
\usepackage{color}
\usepackage{subcaption}
\usepackage{amsmath}
\usepackage[table]{xcolor}
\usepackage{graphicx,epstopdf}
\usepackage{cleveref}
\usepackage{algorithm}
\usepackage{algorithmic}
\usepackage{hyperref}

\def\bfY{\boldsymbol{Y}}
\def\bfD{\boldsymbol{D}}
\def\bfW{\boldsymbol{W}}

\def\bfp{\boldsymbol{\pi}}
\def\bfd{\boldsymbol{d}}
\def\bfw{\boldsymbol{w}}
\def\bfy{\boldsymbol{y}}
\def\bfz{\boldsymbol{z}}
\def\bTheta{\boldsymbol{\Theta}}
\def\btheta{\boldsymbol{\theta}}
\def\mN{\mathcal{N}}
\def\mV{\mathcal{V}}

\Crefformat{figure}{#2Fig.~#1#3}
\Crefmultiformat{figure}{Figs.~#2#1#3}{ and~#2#1#3}{, #2#1#3}{ and~#2#1#3}
\newenvironment{algogo}[1]{
\smallskip
\noindent \hrule\vspace{0.2\baselineskip} \hrule
\begin{small}
\refstepcounter{algo} \center{\bf \textsc{Algorithm \thealgo}}
\\{\center{\bf #1}}
\smallskip
\flushleft
 } {
\end{small}
\smallskip
\hrule\vspace{0.2\baselineskip} \hrule
\smallskip
}
\newcounter{algo}
\renewcommand{\thealgo}{\arabic{algo}}

\begin{document}
%
\title{Fast online 3D reconstruction of dynamic scenes from individual single-photon detection events}
%
%
%

\author{
	Yoann~Altmann,~\IEEEmembership{Member,~IEEE,}
	Stephen~McLaughlin,~\IEEEmembership{Fellow,~IEEE}
	and~Michael~E.~Davies,~\IEEEmembership{Fellow,~IEEE}
	\thanks{}}

\maketitle

\begin{abstract}
In this paper, we present an algorithm for online 3D reconstruction of dynamic scenes using individual times of arrival (ToA) of photons recorded by single-photon detector arrays. One of the main challenges in 3D imaging using single-photon Lidar is the integration time required to build ToA histograms and reconstruct reliable 3D profiles in the presence of non-negligible ambient illumination. This long integration time also prevents the analysis of rapid dynamic scenes using existing techniques. We propose a new method which does not rely on the construction of ToA histograms but allows, for the first time, individual detection events to be processed online, in a parallel manner in different pixels, while accounting for the intrinsic spatiotemporal structure of dynamic scenes. Adopting a Bayesian approach, a Bayesian model is constructed to capture the dynamics of the 3D profile and an approximate inference scheme based on assumed density filtering is proposed, yielding a fast and robust reconstruction algorithm able to process efficiently thousands to millions of frames, as usually recorded using single-photon detectors. The performance of the proposed method, able to process hundreds of frames per second, is assessed using a series of experiments conducted with static and dynamic 3D scenes and the results obtained pave the way to a new family of real-time 3D reconstruction solutions.
\end{abstract}


%
\IEEEpeerreviewmaketitle

\section{Introduction}
\label{sec:intro}
Fast reconstruction of 3D scenes using single-photon light detection and ranging (Lidar) technology is an important challenge which is important in applications such as autonomous driving \cite{hecht2018lidar}, environmental monitoring \cite{mallet2009full,Canutoeaau0137,wallace2014design} and defence \cite{gao2011research}. A growing number of 3D imaging modalities is becoming increasingly popular \cite{Horaud2016}, and single-photon Lidar offers appealing advantages, including low-power, a capability for long-range imaging \cite{Pawlikowska:17} or imaging in complex media such as fog/smoke \cite{Tobin:19} and underwater \cite{maccarone2015underwater,halimiwater} with excellent range resolution (of the order of millimetres \cite{McCarthy:09}). Recently, several algorithms have also been proposed to analyse distributed objects \cite{shin2016computational,tobin2017long,halimi2017,Tachella2019_manipop,Halimi2019SSPD,Tachella2019_ICASSP}, i.e., when multiple surfaces are visible within each pixel. 

Despite pushing the boundaries of 3D reconstruction in extreme environments, single-photon Lidar still suffers from 1) relatively long integration times required to obtain sufficiently reliable data and 2) significant computational requirements to process the resulting large volume of data recorded by single-photon imaging systems. Recent advances in single-photon avalanche diode (SPAD) detector arrays \cite{Ren:18,Henderson2019} have allowed significant reductions in acquisition times over raster scanning systems \cite{McCarthy:09,altmann2016lidar,altmann2016target,halimi2016restoration}, enabling acquisitions with video frame rates. Yet, robust, automated and scalable methods allowing for fast analysis of single-photon data are still required. One of the main bottlenecks of most state-of-the-art 3D reconstruction methods \cite{shin2015photon,altmann2018eusipco,Lindell:2018:3D,rappfew,Tachella2019_manipop,Rapp2019} is that they rely on the construction of histograms of photon times of arrival (ToA) (or batches of detection events), which, when synchronised with a pulsed laser (time correlated single-photon counting, TCSPC) correspond to photon times of flight (ToF), used to infer object ranges. One important exception is the so-called "first-photon" imaging approach \cite{kirmani2014first} whereby the reflectivity and 3D profiles of the scene can be recovered using a single photon per pixel. However, the approach in \cite{kirmani2014first} targets primarily raster scanning Lidar systems, allowing the variable per-pixel acquisition times, i.e., until the first photon is detected. 

In this work, we consider Lidar data acquired using SPAD arrays and investigate a new 3D reconstruction algorithm that does not rely on ToF histograms, but on individual photon detection events. More precisely, we address the problem of 3D reconstruction after each time period (defined in Section \ref{sec:method}) during which each SPAD detector can record at most one detection event. This approach is particularly relevant for applications where the objects in the scene can move significantly faster than the integration period or the number of laser repetitions required to build sufficiently populated ToF histograms. In such cases, the relative movement of the scene with respect to the sensor can produce a 3D blur that produces broader peaks or even multiple returns in some Lidar waveforms, which can jeopardise the 3D reconstruction task. 

Adopting a Bayesian approach, we consider a likelihood model based on the standard single-photon Lidar observation model in the low-flux regime. We then introduce a dynamic model for the spatiotemporal (ST) evolution of the 3D profile. Due to the complex nature of the likelihood (mixture of two distributions) and the structure of the prior model, the standard online estimation methods based on (extended) Kalman filtering \cite{Chui1987} cannot be used directly. As the complexity of the resulting model grows prohibitively with the number of detection events, i.e., over time, we adopt an approximate estimation strategy based on assumed density filtering (ADF) \cite{Lauritzen1992,Boyen1998,minka2001expectation}, whereby the posterior distribution of the 3D profile estimated for a given frame is projected onto a family of more tractable distributions (Gaussian distributions here), which reduces significantly the complexity of the sequential estimation procedure. Although particle filters \cite{Haug2012} could also be considered for approximate inference, this would lead to an increased computational cost induced by the approximation of densities using a large number of particles. It is important to note that the resulting method, which uses each frame (during which at most one photon can be detected per pixel) only once, enables online 3D reconstruction of dynamic scenes with limited memory requirements. Indeed, the individual frames are processed sequentially, resulting in a fix computational cost per frame which is important for any real time implementation. Moreover, thanks to its intrinsically parallel algorithmic architecture, the proposed method is extremely scalable to large arrays and long sequences of frames. Another important advantage is that it does not require the knowledge of the (potentially time varying) ambient illumination level. 

To summarise, the main contributions of this work are:
\begin{itemize}
\item A new Bayesian model for sequential 3D reconstruction using individual photon-detection events
\item An online estimation strategy, proposed to the best of our knowledge for the first time, for reconstruction of dynamic 3D scenes from single-photon data. This method based on assumed density filtering is highly scalable and computationally attractive.
\end{itemize}

The remainder of the paper is organised as follows. Section \ref{sec:method} first recalls the classical observation model for 3D reconstruction using single-photon measurements in the photon-starved regime and describes the Bayesian model and inference strategy proposed for 3D reconstruction using a single frame. The generalisation of this method to online 3D reconstruction of dynamic scenes is detailed in Section \ref{sec:online}. Results of simulations conducted with simulated single-pixel data and sequences of frames are presented and discussed in Section \ref{sec:results} and conclusions are finally reported in Section \ref{sec:conclusion}.

\section{Single frame analysis}
\label{sec:method}
\subsection{Observation model}
\label{subsec:likelihood}
In this work, we consider a sequence of $N$ frames, where each frame of duration $T$ consists of $P$ pixels. This paper addresses the reconstruction of dynamic 3D scenes where each single-photon detector, associated with one pixel, is able to record at most one detection event per pixel and per frame.

Let's first consider an active illumination scenario where the laser emits pulses of light with a repetition/illumination period $T_r=T$.
As described in \cite{rappfew}, assuming that a single surface is visible in each pixel, within each frame $n$, the average photon flux at the detector/pixel $p$ can be modelled as 
\begin{eqnarray}
\label{eq:flux}
\lambda_{p,n}(t) = r_{p,n}s(t - 2d_{p,n}/c) + b_{p,n}, \forall t \in \left[0;T_r\right),
\end{eqnarray}
where $d_{p,n}$ is the instantaneous distance of the object, $c$ is the speed of light in the homogeneous medium between the imaging system and the detector and $r_{p,n}$ is an amplitude parameter related to the reflectivity of the object. Moreover, $b_{p,n}$ represents the ambient illumination and dark count level in the $n$th pixel, which can potentially vary among pixels. Note that $r_{p,n}$ and $b_{p,n}$ also account for the quantum efficiency of the detectors that is not further detailed here for brevity (see \cite{rappfew} for details). Moreover, $s(\cdot)$ is the overall impulse response of the imaging system, which includes the shape of the pulse emitted by the laser and the temporal response of the single-photon detector. As in \cite{rappfew}, we assume that $s(\cdot)$ is known as it can be measured during the calibration of the Lidar system, and that it is well approximated by a Gaussian profile with variance $s^2$. As will be discussed in Section \ref{sec:method}, the proposed method can also be applied when the shape of this impulse response is not Gaussian and changes from one pixel to another due, for instance, to the inhomogeneity of the $P$ detectors.

Over the $n$th illumination period, the detection rate is thus given by 

\begin{eqnarray}
\Lambda_{p,n} = \int_{0}^{T_r} \lambda_{p,n}(t) \textrm{d}t = r_{p,n} S+ B_{p,n}
\end{eqnarray} 
where $B_{p,n}=T_r b_{p,n}$ and where we assume that the object distance is not too close from the minimum $(0)$ and maximum $(T_r c/2)$ admissible ranges such the the integral $S=\int_{0}^{T_r} s(t- 2d_{p,n}/c)  dt $ remains constant, whatever the value of $d_{p,n}$. In the low-flux regime, we have $r_{p,n} (t) S+ B_{p,n}\ll1$, such that the probability of two photons reaching the same detector in a given interval $T_r$ is small and such that the dead-time of the detector can be neglected. In that case, the probability of detection is given by $\pi_{p,n}=1-\exp\left[-\Lambda_{p,n}\right] \approx \Lambda_{p,n}$ and the probability of a detected photon being associated with the original emitted pulse, denoted by $w_{p,n}$ is given by $w_{p,n}=(r_{p,n}S)/\Lambda_{p,n}$. Let $z_{p,n} \in (0;1)$ be a binary label indicating a detection event (i.e., when $z_{p,n}=1$) in pixel $p$ for the $n$th frame, such that 
\begin{eqnarray}
\label{eq:detect_prob}
f(z_{p,n}=1|\pi_{p,n})=\pi_{p,n}. 
\end{eqnarray}
When $z_{p,n}=1$, the observation model for the measured time of arrival $y_{p,n} \in \left[0;T_r\right)$ in pixel $p$ and frame $n$ can be expressed as\\ 
$f(y_{p,n}  | z_{p,n} =1, w_{p,n}, d_{p,n} )$
\begin{eqnarray}
\label{eq:lik_single}
 = w_{p,n} f_s\left(y_{p,n} - \frac{2 d_{p,n}}{c}\right) + (1-w_{p,n}) \mathcal{U}_{\left[0;T_r\right)} (y_{p,n} ),
\end{eqnarray}
where $\mathcal{U}_{\left[0;T_r\right)} (\cdot)$ is the uniform distribution defined on $\left[0;T_r\right)$, and with $f_s(t - 2d_{p,n}/c)=S^{-1}s(t - 2d_{p,n}/c), \forall (p,n)$. Moreover, we use the notation $y_{p,n} = \emptyset$ when no detections were recorded in the $p$th pixel within the $n$th frame.
 
Assume now that a frame lasts $N_r$ laser repetition periods, i.e., $T=N_rT_r$ and that the detector is only able to record at most one detection event during that frame. If the observation conditions have not changed during the $N_r$ repetitions, the probability of detection is given by $\tilde{\pi}_{p,n}=1-\exp\left[-N_r\Lambda_{p,n}\right]$. However, in the low-flux regime, Eq. \eqref{eq:lik_single} still applies. Consequently, although each frame can result from more than one illumination periods, the observation models \eqref{eq:detect_prob} and \eqref{eq:lik_single} are still valid by replacing $\pi_{p,n}$ by $\tilde{\pi}_{p,n}$ in \eqref{eq:detect_prob}, provided that the observation conditions have not changed over the period $T$. This observation can be useful for practical applications since in the low-flux regime, imposing $r_{p,n} S+ B_{p,n}\ll1$ (for each $T_r$ interval) leads to extremely sparse detection events and large volumes with $T=T_r$, while using $T=N_rT_r$ (for a given $T_r$) allows both reduced data volume and higher per-frame detection rates.

In this paper, we address the problem of estimating $\bfD=\{d_{p,n}\}_{p,n}$ from the set of observations $\bfY=\{y_{p,n}\}_{p,n}$. 
As discussed in the introduction of this paper, although it is possible to develop batch-based methods for recovering $\bfD$ given all the $NP$ observations \cite{altmann2016lidar,rappfew}, such approaches can become computationally prohibitive for large numbers of pixels, but more importantly for long temporal sequences. Moreover, these existing approaches do not specifically deal with time varying scenes, and do not use a spatiotemporal models. While the method recently proposed by Halimi et al. \cite{Halimi2019SSPD,Tobin:19} can be used for varying scenes (sequences of batches), it is not designed to handle long temporal sequences either as all the batches are proposed simultaneously. Indeed, the ADMM-based method developed in \cite{Halimi2019SSPD} seems more suited for multispectral measurements. Thus, here we adopt a sequential approach where the $N$ frames are processed one by one and only once, allowing for fast estimation and reduced memory requirements. In the remainder of the paper, we thus use \eqref{eq:detect_prob}-\eqref{eq:lik_single} as our observation model.

The next paragraph introduces the Bayesian model and estimation strategy used to process a single frame, assuming that $\bfW=\{w_{p,n}\}_{p,n}$ is known. The generalisation of the proposed updated to online 3D reconstruction, including the sequential estimation of  $\bfW$ will be discussed in Section \ref{sec:online}.

\subsection{Estimation strategy}
\label{subsec:estim_singleframe}
As mentioned above, we first investigate the estimation of $\bfd_n=\{d_{p,n}\}_{p}$ from a set of measurements $\bfy_n=\{y_{p,n}\}_{p}$ associated with the $n$th frame. 
Assuming the detection events in different pixels are mutually independent (given the other parameters in \eqref{eq:lik_single}), the joint likelihood can be expressed as 
\begin{eqnarray}
\label{eq:joint_lik}
f(\bfy_n  | \bfz_n,\bfw_n,\bfd_n) = \prod_{p=1}^P  f(y_{p,n}  | z_{p,n}, w_{p,n}, d_{p,n} ),
\end{eqnarray}
with $\bfz_n=\{z_{p,n}\}_{p}$, $\bfw_n=\{w_{p,n}\}_{p}$ and $f(y_{p,n}=\emptyset | z_{p,n}=0, w_{p,n}, d_{p,n} )=1$.

To obtain a tractable and computationally efficient ADF-based estimation strategy, we propose to define independent prior distributions for the target ranges in a given frame, i.e., $f(\bfd_n|\bTheta_n)=\prod_{p=1}^P f(d_{p,n}|\btheta_{p,n})$. Despite the apparent lack of prior correlation between the elements of $\bfd_n$ (given the set of parameters in $\bTheta_n=\{\btheta_{p,n}\}_p$), it is possible to enforce ST correlations by defining $\bTheta_n$ using $\bfd_{n-1}$, as will be discussed in Section \ref{sec:online}. For now, let's assume that each distance $d_{p,n}$ is assigned a fully specified mixture of $M$ Gaussian distributions as follows
\begin{eqnarray}
\label{eq:indiv_prior}
f(d_{p,n}|\btheta_{p,n}) \sim \sum_{m=1}^M u_{p,n}^{(m)} \mN(d_{p,n};\mu_{p,n}^{(m)},\sigma_{p,n}^{2(m)}), 
\end{eqnarray}
where $\mN(\cdot;\mu_{p,n}^{(m)},\sigma_{p,n}^{2(m)})$ is a Gaussian distribution with mean $\mu_{p,n}^{(m)}$ and variance $\sigma_{p,n}^{2(m)}$ and where  $\btheta_{p,n}=\left\lbrace \mu_{p,n}^{(m)},\sigma_{p,n}^{2(m)}\right\rbrace_m$. The weights $\{u_{p,n}^{(m)}\}_m$ in \eqref{eq:indiv_prior} satisfy $\sum_{m=1}^M u_{p,n}^{(m)} =1, \forall (p,n)$ and their value, as well as that of $M$ will be discussed in Section \ref{sec:online}. Since the joint likelihood \eqref{eq:joint_lik} and the joint prior distribution \eqref{eq:indiv_prior} can be factorised over the $P$ pixels, the resulting posterior distribution 
given by \\
$f(\bfd_n|\bfy_n, \bfz_n, \bfw_n,\btheta_{p,n})$
\begin{eqnarray}
\label{eq:joint_post}
 & \propto &  f(\bfy_n  | \bfz_n, \bfw_n,\bfd_n) f(\bfd_n|\bTheta_n)) \nonumber\\
 & \propto & \prod_{p=1}^P  f(y_{p,n}  | z_{p,n}, w_{p,n}, d_{p,n} ) f(d_{p,n}|\btheta_{p,n}) , 
\end{eqnarray}
can also be factorised over the $P$ pixels and the $P$ range parameters in $\bfd_n$ can thus be estimated independently, in a parallel manner. Consequently, we simply summarise the update for one parameter $d_{p,n}$, i.e., for pixel $p$. 
If $z_{p,n}=0$, $d_{p,n}$ does not appear in the data likelihood. In that case, the posterior distribution of $d_{p,n}$ reduces to its prior \eqref{eq:indiv_prior}. If $z_{p,n}=1$, the posterior distribution of $d_{p,n}$ is the following mixture\\
$f(d_{p,n}| y_{p,n}, z_{p,n}=1, w_{p,n}, \btheta_{p,n})$
\begin{eqnarray}
\label{eq:post_d}
 \propto f(d_{p,n}|\btheta_{p,n}) f(y_{p,n}  | z_{p,n} =1, w_{p,n}, d_{p,n} ), 
\end{eqnarray}
which is a mixture of $2M$ Gaussian distributions when $f_s(\cdot)$ is also Gaussian. Although $f(d_{p,n}| y_{p,n}, z_{p,n}=1,w_{p,n}, \btheta_{p,n})$ seems to be only known up to a multiplicative constant, its normalising constant, as well as the mixture weights and the mean/variances of each component of the mixture can be computed analytically by integrating \eqref{eq:post_d} with respect to (w.r.t.) $d_{p,n}$. The moments of $f(d_{p,n}| y_{p,n}, z_{p,n}=1, w_{p,n}, \btheta_{p,n})$, and in particular its mean and variance can then be computed as for any mixture of distributions \cite[Chap. 1]{frühwirth2006}. These summary statistics are then used to obtain a point estimate (i.e., the mean) of $d_{p,n}$, as well as corresponding measures of uncertainty (through the variance). 
When $f_s(\cdot)$ is not Gaussian, it is in general not possible to compute analytically the mean and variance of $f(d_{p,n}| y_{p,n}, z_{p,n}=1, w_{p,n}, \btheta_{p,n})$, but it is possible to resort to numerical integration tools \cite{Wand2011,Perelli2016} such as Gaussian quadrature or Laplace approximation to approximate the integrals 
\begin{eqnarray}
\int f_s(y_{p,n}- 2d_{p,n}/c)\mN(d_{p,n};\mu_{p,n}^{(m)},\sigma_{p,n}^{2(m)}) \textrm{d} d_{p,n},
\end{eqnarray}
and in turn the moments of $f(d_{p,n}| y_{p,n}, z_{p,n}=1, w_{p,n}, \btheta_{p,n})$.

\section{Online estimation}
\label{sec:online}
\subsection{Approximation using Assumed Density Filtering}
Estimating the posterior mean and variance of $d_{p,n}$ presents a great advantage for online estimation, beyond simply providing summary statistics about the current range profile. It allows, by propagating simply the first and second-order moments of the current posterior distributions, the use of a tractable adaptive estimation procedure. Indeed, if the prior distribution of $d_{p,n}$ consists of $M$ components (as in \eqref{eq:indiv_prior}), its posterior will contain $2M$ components (only $M$ if $z_{p,n}=0$) and if a classical Gaussian random walk is then used to model $f(d_{p,n+1}|d_{p,n})$, the posterior distribution of  $d_{p,n+1}$ will present $4M$ terms after marginalisation of $d_{p,n}$. That number will thus increase prohibitively as $n$ increases. 
The basic principle of assumed density filtering in this case is to approximate $f(d_{p,n}| y_{p,n}, w_{p,n}, \btheta_{p,n})$ by a more tractable distribution that can then be used to build a new prior distribution for $d_{p,n+1}$. 
While it is possible to construct complex approximations of \eqref{eq:post_d} using a fixed (reduced) number of Gaussian components, here we simply use an approximation based on a single Gaussian $q(d_{p,n})$. In a similar fashion to classical assumed density filtering \cite{Lauritzen1992,Boyen1998} and expectation-propagation\cite{minka2001expectation}, this approximation is found by minimising the following Kullback-Leibler divergence
\begin{eqnarray}
\label{eq:KL_div}
KL \left[f(d_{p,n}| y_{p,n}, w_{p,n}, \btheta_{p,n}) || q_{p,n}(d_{p,n})\right]
\end{eqnarray} 
w.r.t. $q_{p,n}(d_{p,n})$ which belongs to the family of Gaussian distributions. This minimisation reduces to matching the mean and variance of $f(d_{p,n}| y_{p,n}, w_{p,n}, \btheta_{p,n})$ and $q_{p,n}(d_{p,n})$, hence the discussion about the estimation of the moments of $f(d_{p,n}| y_{p,n}, w_{p,n}, \btheta_{p,n})$ in Section \ref{subsec:estim_singleframe}. 

\subsection{Spatiotemporal dynamic model for the range profile}
A classical choice for modelling relatively slowly evolving parameters relies on (Gaussian) random walks. Whilst this approach is easy to implement, it does not allow, using simply $f(d_{p,n+1}|d_{p,n}), \forall p$, for rapid changes as might occur when the imaging system or the scene moves orthogonally to the direction of observation, whereby a foreground object can disappear from one pixel and appear in neighbouring pixels. To alleviate issues associated with such changes while keeping the estimation strategy tractable, we define, for each pixel, a local neighbourhood $\mV_p$ of $M$ neighbours (including the current pixel) and define the following prior model \\
$f(d_{p,n+1}|\btheta_{p,n+1}) $
\begin{eqnarray}
\label{eq:predict}
\propto \sum_{p' \in \mV_p} \nu_{p'} f_{\gamma^2}(d_{p,n+1}|d_{p',n})q_{p',n}(d_{p',n}),
\end{eqnarray}
where $\{q(d_{p,n})\}_{p,n}$ are the Gaussian approximating posterior distributions of $\{d_{p,n}\}_{p,n} $ computed by minimising \eqref{eq:KL_div} and $f_{\gamma^2}(\cdot | d_{p,n'}) = \mN(\cdot;d_{p,n'},\gamma^2)$ is a Gaussian random walk which models (through its variance $\gamma^2$) the expected amount of movement of the objects of the scenes along the direction of observation, between two frames. In \eqref{eq:predict}, the weights $\nu_{p'}$ are chosen such that $\nu_{p'=p}=\nu$ and  $\nu_{p'\neq p}=(1-\nu)/(M-1)$, i.e., $\nu$ represents the probability of a surface to remain in the same pixel in the next frame. While we choose in \eqref{eq:predict} a relatively simple ST model based a random walk, more complex local dynamic models (e.g., using object velocity) could also be used. As long as the prior model reduces to a mixture of Gaussian distributions, the proposed estimation strategy can still be used. 
 
In a similar fashion to \eqref{eq:post_d}, Eq. \eqref{eq:predict} is a finite mixture a $M$ Gaussian distributions whose weights, and individual means and variances, gathered in $\btheta_{p,n+1}$ can be computed analytically by integration of the right-hand side of $\eqref{eq:predict} $ w.r.t. $\{d_{p',n}\}_{p' \in \mV_p}$. Using this strategy, the number of components of $f(d_{p,n+1}|\btheta_{p,n+1})$ remains the same as for $f(d_{p,n}|\btheta_{p,n})$, that is, $M$. The size of the neighbourhood can be adjusted using prior information about the scene on interest: small neighbourhoods  (e.g., $M=5$ using $4$ neighbouring pixels, as used in this work) will be sufficient for slowly moving scenes but larger sets of pixels might be needed if objects can move by several pixels in the image plane. 

\subsection{Estimation of the other model parameters}

\begin{algogo}{O3DSP algorithm}
\label{algo:algo1}
 \begin{algorithmic}[1]
			\STATE \underline{Fixed input parameters:} Variance of RW for dynamic model: $\gamma^2$, Neighbourhood size $M$, temporal smoothing parameter $\alpha$, parameter of GMM $\nu$.
			\STATE \underline{Initialization ($n=0$)}
			\STATE Set $(\bar{w}_{p,0}, q_{p,0}), \forall p$ 
			\FOR{$n=1,\ldots N$}
				\FOR{$p=1,\ldots P$}
				\STATE Compute prior model $f(d_{p,n}|\btheta_{p,n}) $ from \eqref{eq:predict}. 
				\STATE Compute exact posterior distribution $f(d_{p,n}| y_{p,n}, \bar{w}_{p,n}, \btheta_{p,n})$ in \eqref{eq:post_d}.
				\STATE Compute $q_{p,n}(d_{p,n})$ using \eqref{eq:KL_div}.	
				\ENDFOR	
				\STATE Compute $\bar{\bfw}_{n+1} = (1-\alpha)\bar{\bfw}_{n} + \alpha \hat{\bfw}_{n}$.
				\STATE Optional: Apply smoothing operator to $\bar{\bfw}_{n+1}$.	
			\ENDFOR
		\STATE Set 
\end{algorithmic}
\end{algogo} 

Interestingly, the proposed 3D reconstruction method does not rely on the knowledge of the detection probabilities $\{\bfp_n\}_n$ since they do not intervene in the estimation of $\bfd_n$ which only relies on $\{\bfy_n\}_n$. In particular, this method does require knowledge of the number $N_r$ of illumination periods during each frame, which is used in the probabilities of detection $\{\tilde{\pi}_{p,n}\}_{p,n}$ (see discussion below Eq. \eqref{eq:lik_single}). Thus, the only important and generally unknown parameters are the probabilities of signal detection events in $\bfW$. In a similar fashion to the approach we proposed for $\bfD$, the elements of $\bfW$ can be included in a Bayesian model and assigned iteratively prior distributions for online estimation, i.e., by computing the posterior distribution of $(\bfd_n, \bfw_n)$ at each frame, and by approximating this distribution to build a tractable prior distribution $f(\bfd_{n+1}, \bfw_{n+1}| \bfd_n, \bfw_n)$. However, this is not the approach we adopt here as it makes the estimation procedure more computationally demanding, in particular when computing the marginal moments, or more generally expectations w.r.t the posterior distribution of $(\bfd_n, \bfw_n)$ during the KL divergence minimisation. Instead, we use the following simple heuristic method which provides satisfactory results in practice. Let $\bar{\bfw}_n$ be an estimate of $\bfw_n$ obtained from the previously observed data $\{\bfy_n\}_{n=1,\ldots,n-1}$. Our aim here is to propose an estimate $\bar{\bfw}_{n+1}$ of $\bfw_{n+1}$, which depends on $\bar{\bfw}_n$ and the data $\bfy_n$. We first define an instantaneous estimator $\hat{\bfw}_n=\{\hat{w}_{p,n}\}_p$ with $\hat{w}_{p,n}=\bar{w}_{p,n}$ if $y_{p,n} = \emptyset$. If $y_{p,n} \neq \emptyset$, $\hat{w}_{p,n}$ is obtained from \eqref{eq:post_d} where $\bfw_n$ has been replaced by $\bar{\bfw}_n$. More precisely,  $\hat{w}_{p,n}$ is the posterior probability of the current detection event to be a signal detection. Thus $\hat{w}_{p,n}$ can be obtained by summing the weights of the $M$ components involving $f_s(\cdot)$ in \eqref{eq:post_d}. The updated vector of probabilities is obtained using $\bar{\bfw}_{n+1} = (1-\alpha)\bar{\bfw}_{n} + \alpha \hat{\bfw}_{n}$, where $\alpha \in (0;1)$ is an attenuation parameter to be tuned depending on the expected variations of $\bfw_n$ over time. Note that it is also possible to apply a smoothing post-processing step, e.g., standard gaussian filtering to $\bar{\bfw}_{n+1}$ to further refine the estimate of $\bfw_{n+1}$ since these parameters are often expected to be spatially correlated in each frame. As mentioned above, this strategy is simple and does not significantly degrade the performance of the 3D reconstruction method in most scenarios. The pseudo-code of the proposed method, referred to as O3DSP (for Online 3D reconstruction using Single-Photon data) is presented in Algo. \ref{algo:algo1}. 

Another important issue that might arise is the occurrence of a new object in the field of view. A particularly challenging scenario is the appearance of an object initially occluded by another object. In such cases, it is possible to add an extra component  in \eqref{eq:predict}, e.g., whose mean and variance can be related to the mean/median and dispersion of the  $\{d_{p',n}\}_{p'}$, respectively. This approach would be efficient to capture new objects appearing between a foreground object and the background. However, as will be shown in Section \ref{sec:results}, such an extra term does not seem necessary as the proposed ST model naturally enforces large variances around edges, which in turn allows initially occluded objects to be detected.  Note that more complex and principled strategies should be developed to handle more challenging occlusion scenarios and situations where pixels do not contain any objects, which are out of scope of this work. This point will be discussed in the conclusion of this study.  
New objects can also enter the field of view from any side. To address this problem, we include, for the pixels around the edges of the image, and additional Gaussian component (with a large variance) in the mixture \eqref{eq:predict} such that the resulting prior allows at the same time, ranges similar to those in nearby pixels but also significantly different ranges induced by the presence of new objects.

Finally, the proposed algorithm can also be applied in the presence of faulty pixels for which $\pi_{p,n}=0$. For these pixels, the range information will be inferred using the inpainting capability of the model in \eqref{eq:predict}.

\section{Results}
\label{sec:results}
In this section, we discuss the performance of O3DSP through a series of experiments conducted with simulated data whereby ground truth is available for comparison. We first investigate the main parameters influencing the reconstruction performance using individual pixels, i.e., without accounting for information provided by neighbouring pixels. Then, we investigate the reconstruction of static and dynamic scenes using photon-starved measurements.
\subsection{Single-pixel experiments}
In Sections \ref{sec:method} and \ref{sec:online}, we have assumed that the measured times of arrivals follow continuous distributions, i.e., they are either uniformly distributed over $\left[0;T_r\right)$ or Gaussian distributed. However, SPAD detectors have a finite timing resolution, whereby the measured times of arrival follow discrete distributions defined on a finite support. Fortunately, state-of-the-art SPADs \cite{altmann2016lidar,Ren:18,Henderson2019} present a timing resolution which is much smaller that the support of $f_s(\cdot)$ and thus than $T_r$. Consequently, assuming continuous measurements does not significantly bias the estimation performance. Should the temporal resolution of the SPADs be coarser, O3DSP can still be applied using dither on the discrete measured ToAs \cite{Rapp2018_dither}.

\begin{figure}[ht!]
	\center
	\includegraphics[width=\columnwidth]{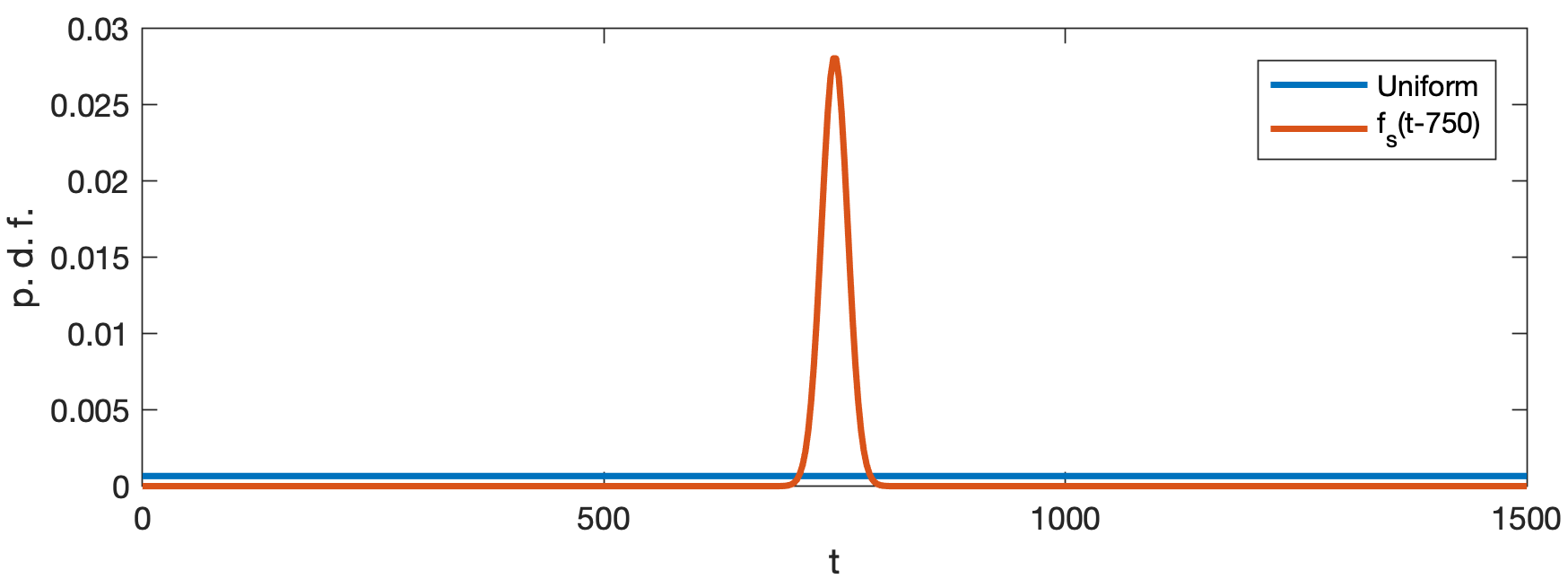}\\
	\caption{Probability density functions (p.d.f.) of the time of arrival of a signal photon (red) for $d=750$ and a background photon (blue), for $T_r=1500$ and $s^2=200$.}
	\label{fig:IRFs}
\end{figure}

In all the simulation results presented in this paper, we use the arbitrary illumination period $T_r=1500$ (unless stated otherwise) and $f_s(\cdot)$ is modelled by a Gaussian distribution with variance $s^2=200$ and without loss of generality, with use $c/2=1$. The distributions of the times of arrival of signal and background photons for $d=750$ are depicted in Fig. \ref{fig:IRFs}. To initialise the algorithm, we used $\bar{w}_{p,0}=0.5, \forall p$ and the Gaussian initial approximations $q_{p,0}(\cdot), \forall p$ are set identically such that their mean is $T_r/2$ and their variance allows the entire interval $(1,T_r)$ to be in the high probability region. This leads to a weakly informative initialisation that we use to assess the convergence of the algorithm. As will be discussed below, more efficient initialisations can also be used.

\begin{figure}[ht!]
	\center
	\includegraphics[width=\columnwidth]{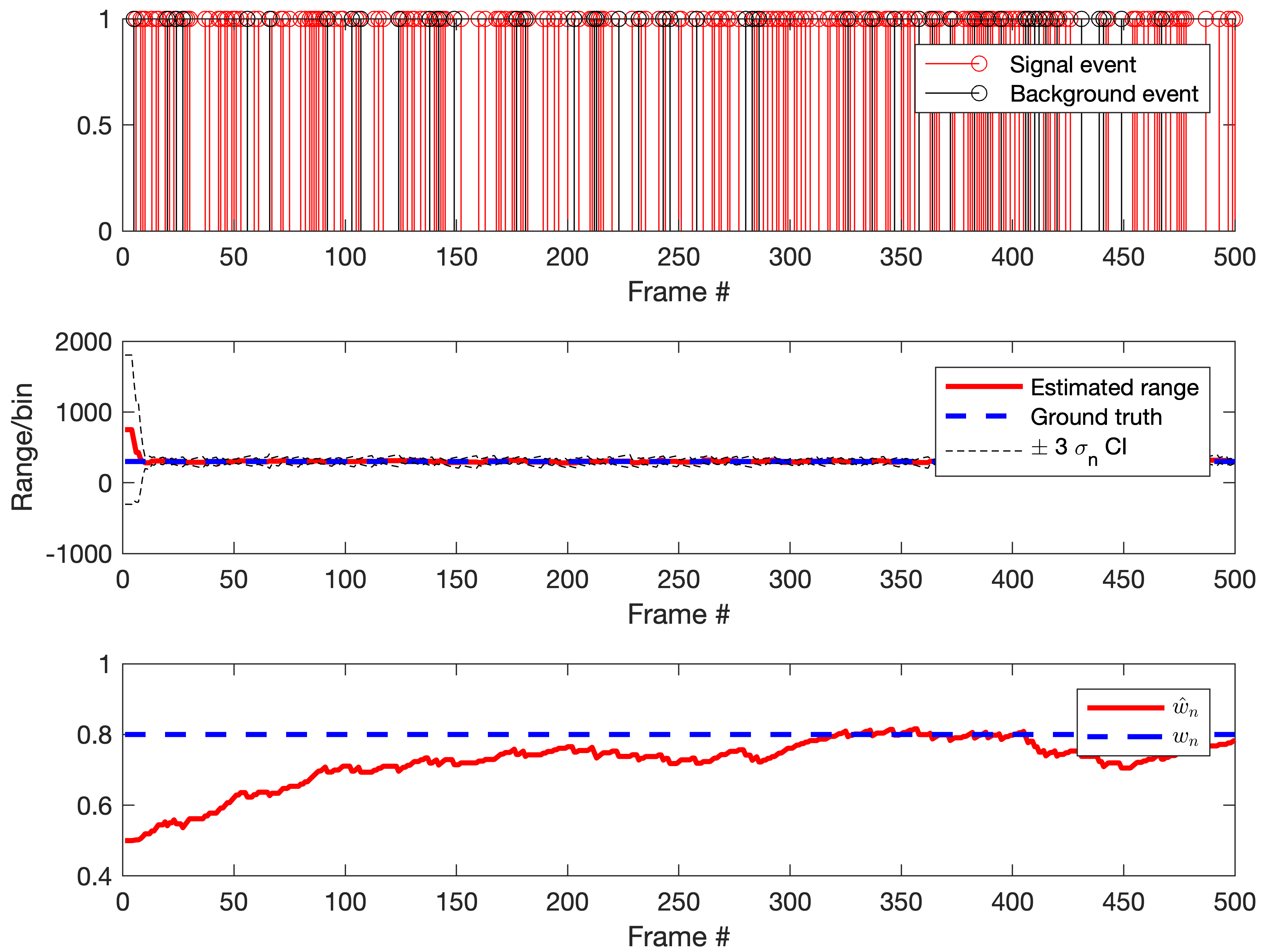}\\
	\vspace{-0.3cm}
	\caption{Top: Examples of background (black) and signal (red) detection events for $N=500$,  $\pi_{p,n}=0.5$, $w_{p,n}=0.8$. Middle: Estimation of $\{d_{p,n}\}_n$ for $\alpha=0.01$ and $\gamma^2=100$. Bottom: online estimates $\{\bar{w}_{p,n}\}_n$ (red lines) of $w_{p,n}$ for $\alpha=0.01$ and $\gamma^2=100$.}
	\label{fig:single_pix1}
\end{figure}

\begin{figure}[ht!]
	\center
	\includegraphics[width=\columnwidth]{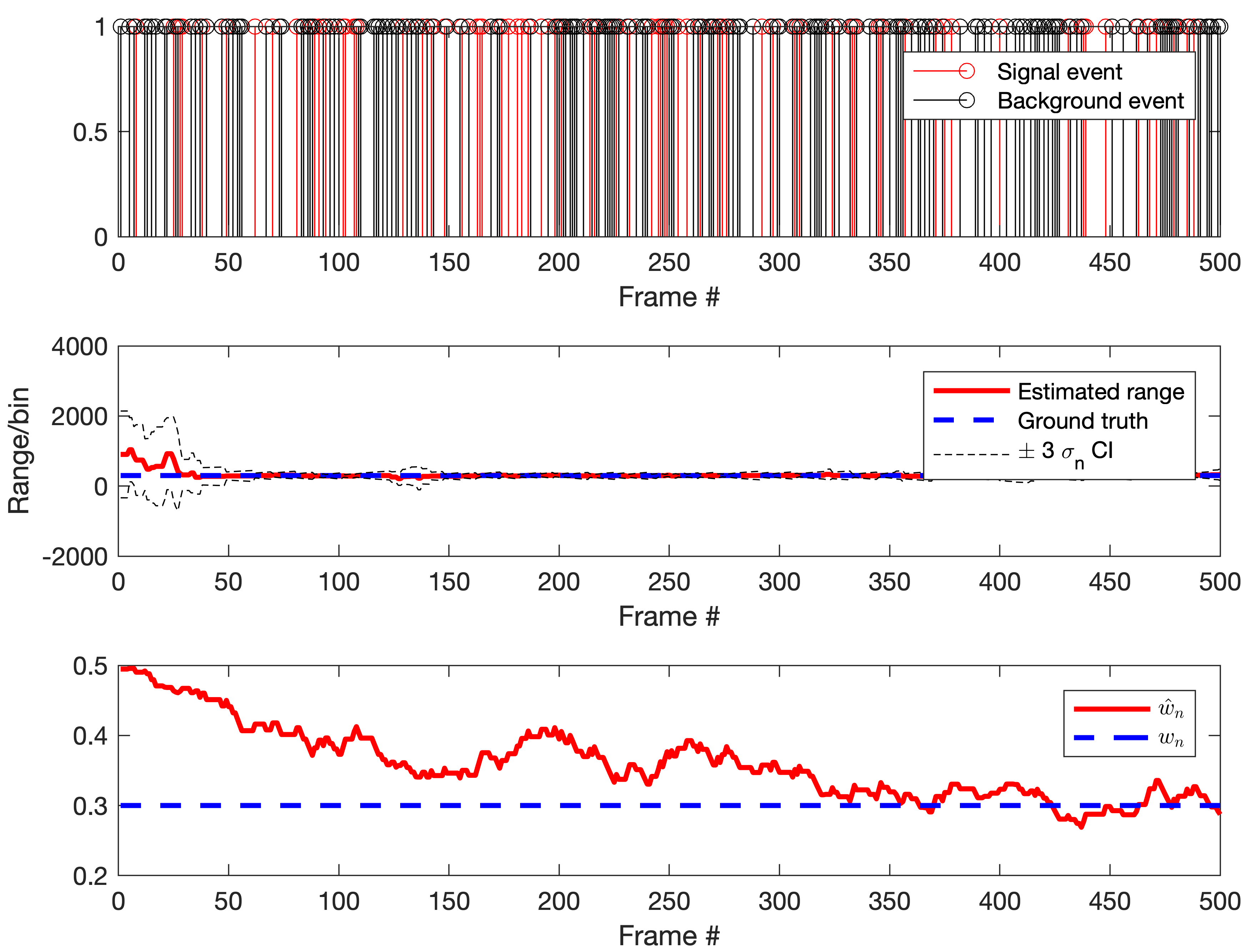}\\
	\vspace{-0.3cm}
	\caption{Top: Examples of background (black) and signal (red) detection events for $N=500$,  $\pi_{p,n}=0.5$, $w_{p,n}=0.3$. Middle: Estimation of $\{d_{p,n}\}_n$ for $\alpha=0.01$ and $\gamma^2=100$. Bottom: online estimates $\{\bar{w}_{p,n}\}_n$ (red lines) of $w_{p,n}$ for $\alpha=0.01$ and $\gamma^2=100$.}
	\label{fig:single_pix2}
\end{figure}

First, we investigate, the impact of $w_{p,n}$ on the estimation of $d_{p,n}$ for a given probability of detection $\pi_{p,n}$. Here the number of frames is set to $N=500$,  $\pi_{p,n}=0.5$, $d_{p,n}=300$ and $\alpha=0.01$. Figs. \ref{fig:single_pix1} and \ref{fig:single_pix2}, compare the convergence of $\{d_{p,n}\}_n$ and $\{\bar{w}_{p,n}\}_n$ for $w_{p,n}=0.8$ (Fig. \ref{fig:single_pix1}) and $w_{p,n}=0.3$ (Fig. \ref{fig:single_pix2}). The top subplots depict the frames during which background (in black) and signal (in red) detections are recorded. The middle subplots depict the mean (red lines) and $\pm 3$ standard deviation intervals (black dashed lines) obtained by minimising \eqref{eq:KL_div}. The bottom subplots represent the online estimates $\{\bar{w}_{p,n}\}_n$ (red lines) of $w_{p,n}$. As can be seen from Figs. \ref{fig:single_pix1} and \ref{fig:single_pix2}, the estimate of $d_{p,n}$ converges faster with $w_{p,n}=0.8$ than with $w_{p,n}=0.3$ (faster convergence to the ground truth and smaller uncertainty). This phenomenon is to be expected as the number of signal detections increases with $w_{p,n}$, which in turn increases the amount of information about $d_{p,n}$. On the other hand, the convergence of $\bar{w}_{p,n}$ seems similar in both cases (around 200-300 frames).

\begin{figure}[ht!]
	\center
	\includegraphics[width=\columnwidth]{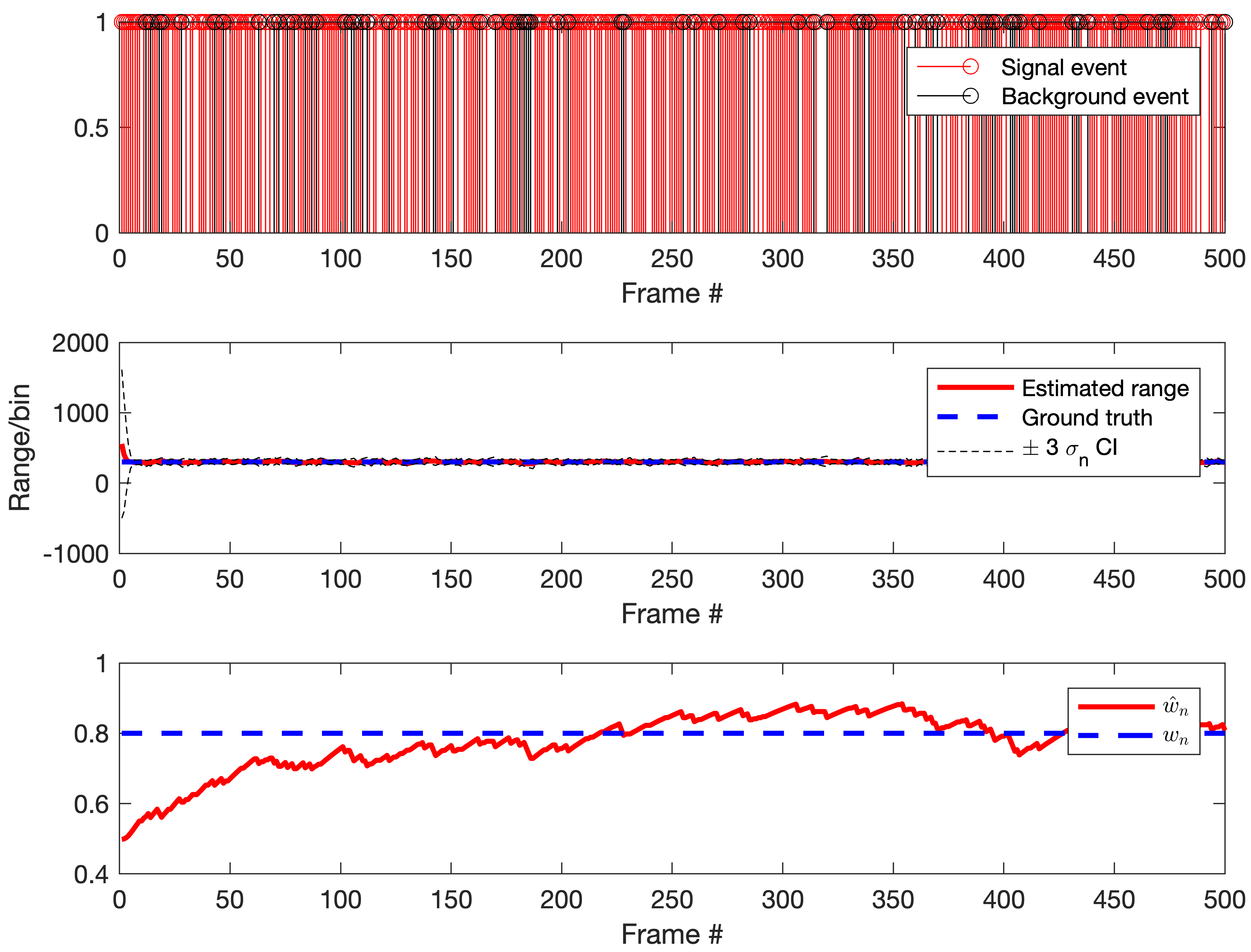}\\
	\vspace{-0.3cm}
	\caption{Top: Examples of background (black) and signal (red) detection events for $N=500$,  $\pi_{p,n}=0.8$, $w_{p,n}=0.3$. Middle: Estimation of $\{d_{p,n}\}_n$ for $\alpha=0.01$ and $\gamma^2=100$. Bottom: online estimates $\{\bar{w}_{p,n}\}_n$ (red lines) of $w_{p,n}$ for $\alpha=0.01$ and $\gamma^2=100$.}
	\label{fig:single_pix3}
\end{figure}

Fig. \ref{fig:single_pix3} shows the estimation of $d_{p,n}$ and $w_{p,n}$ with $\pi_{p,n}=0.8$, $d_{p,n}=300$ and $w_{p,n}=0.3$. As expected, the convergence of $\{d_{p,n}\}_n$ is faster than in Fig. \ref{fig:single_pix2} since its estimation is directly related to the number of signal detections which increases with $\pi_{p,n}$ (for a fixed $w_{p,n}$).

\begin{figure}[ht!]
	\center
	\includegraphics[width=\columnwidth]{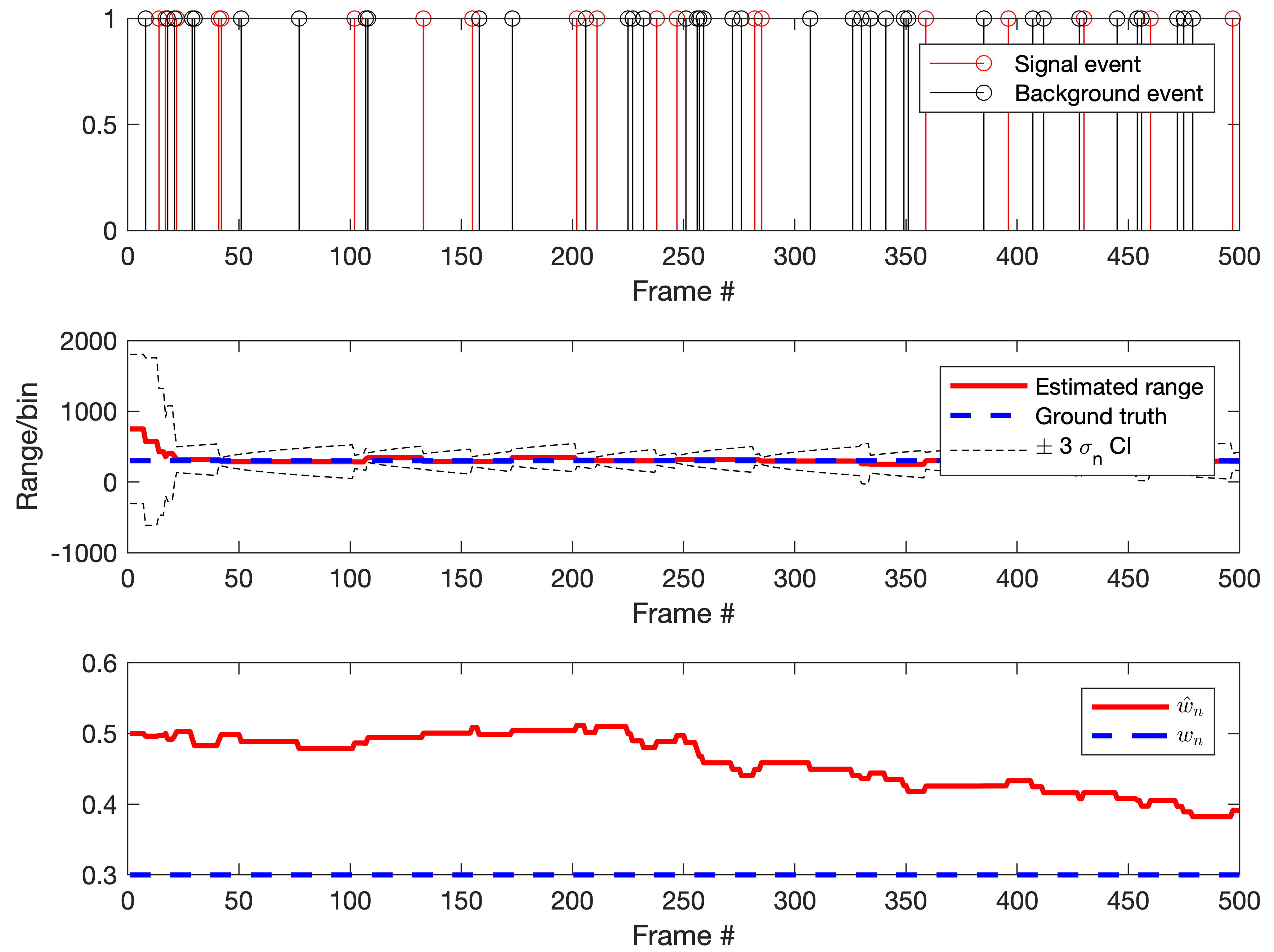}\\
	\vspace{-0.3cm}
	\caption{Top: Examples of background (black) and signal (red) detection events for $N=500$,  $\pi_{p,n}=0.8$, $w_{p,n}=0.3$. Middle: Estimation of $\{d_{p,n}\}_n$ for $\alpha=0.01$ and $\gamma^2=100$. Bottom: online estimates $\{\bar{w}_{p,n}\}_n$ (red lines) of $w_{p,n}$ for $\alpha=0.01$ and $\gamma^2=100$.}
	\label{fig:single_pix4}
\end{figure}

Reducing the probability of detection has an impact on the estimation of $d_{p,n}$ and $w_{p,n}$, as can be seen in Fig. \ref{fig:single_pix4}, where $\pi_{p,n}=0.1$ and $w_{p,n}=0.3$. In this case, with an average of $50$ detection events for $N=500$ frames ($30\%$ of which being signal detections), the convergence speed of $\{\bar{w}_{p,n}\}_n$ is reduced and the uncertainty about $d_{p,n}$ increases due to the lack of information provided by the data. In such difficult scenarios, the proposed method might not converge toward the correct solution without using additional information, which is a well known potential limitation of ADF \cite{minka2001expectation}. However, as will be shown in Section \ref{sec:subsec:videos}, the proposed ST model using information contained in neighbouring pixels (see \eqref{eq:predict}) yields satisfactory results in the photon-starved regimes considered here. 

\begin{figure}[ht!]
	\center
	\includegraphics[width=\columnwidth]{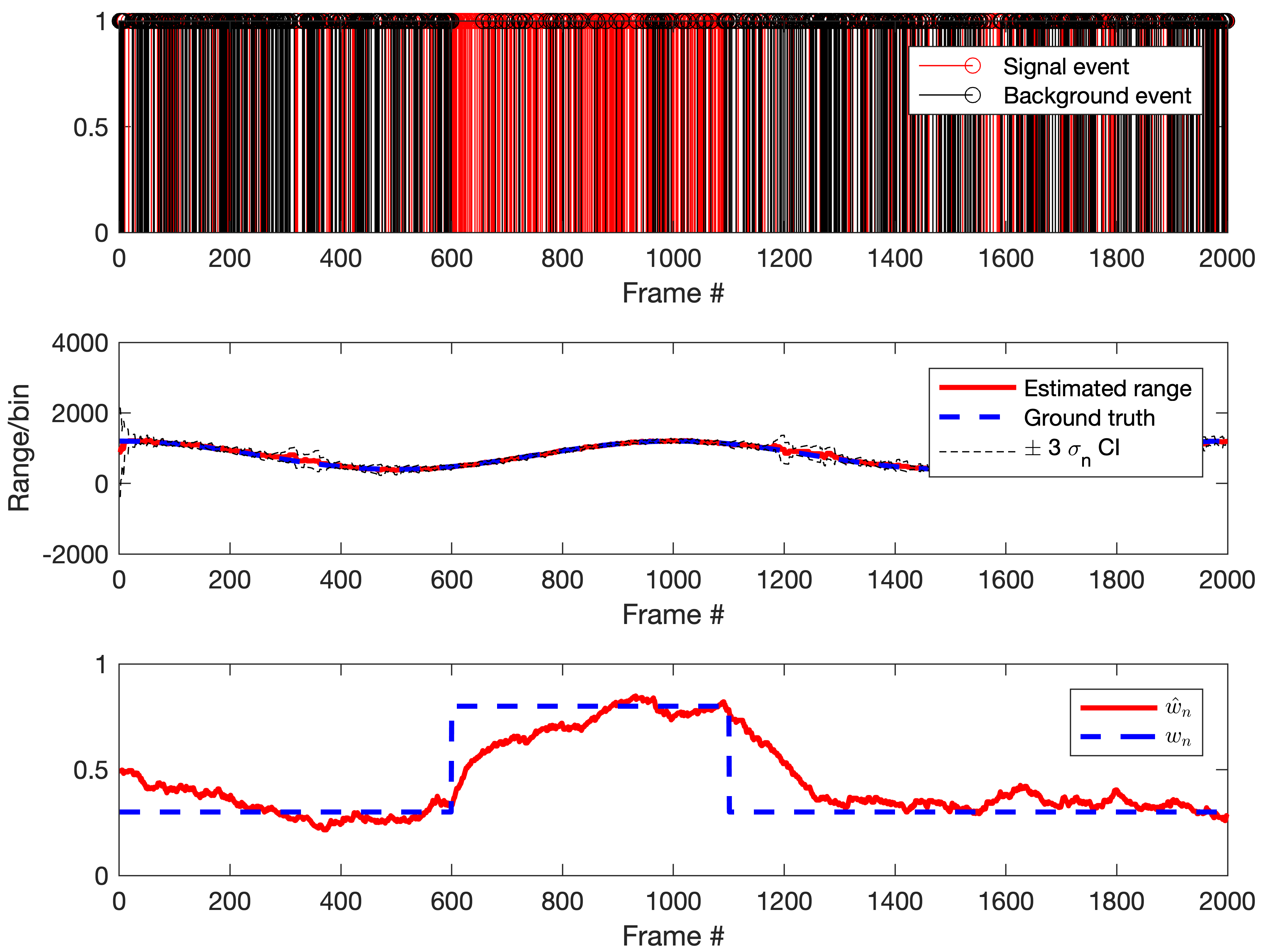}\\
	\vspace{-0.3cm}
	\caption{Analysis of dynamic scene (single pixel) with smooth changes of $d_{p,n}$ and sudden changes of $w_{p,n}$. Top: Examples of background (black) and signal (red) detection events for $N=2000$,  $\pi_{p,n}=0.5$. Middle: Estimation of $\{d_{p,n}\}_n$ for $\alpha=0.01$ and $\gamma^2=100$. Bottom: online estimates $\{\bar{w}_{p,n}\}_n$ (red lines) of $w_{p,n}$.}
	\label{fig:single_pix5}
\end{figure}

We also evaluate the performance of O3DSP by analysing a single-pixel measurement where the object range describes a sine wave and where $w_{p,n}$ experiences two sudden changes (see Fig. \ref{fig:single_pix5}). This figure has been obtained with $\pi_{p,n}=0.5$. As can be seen in the top and bottom subplots of Fig. \ref{fig:single_pix5}, the probability of signal detection $w_{p,n}$ is changed successively from $w_{p,n}=0.3$ to $w_{p,n}=0.8$ and back to $w_{p,n}=0.3$. This figure shows that O3DSP is able to satisfactorily track the changes of $d_{p,n}$ without noticeable delay and that about $200-300$ frames are required for $\bar{w}_{p,n}$ to converge around the correct value. 

\begin{figure}[ht!]
	\center
	\includegraphics[width=\columnwidth]{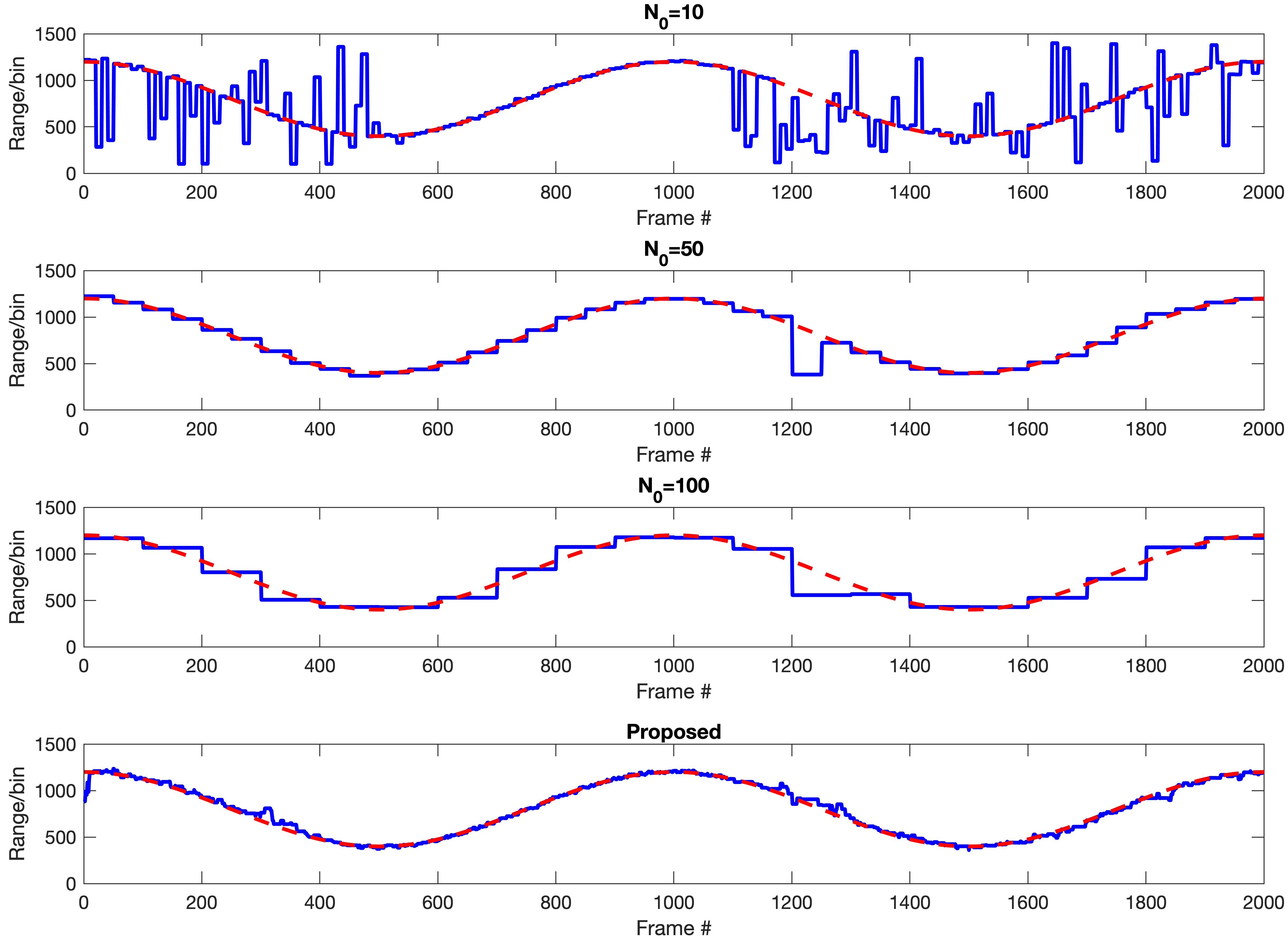}\\
	\vspace{-0.3cm}
	\caption{Three top plots: depth estimates obtained via cross-correlation for batches of $N_0=10$, $N_0=50$ and $N_0=100$. Bottom: Estimation of $\{d_{p,n}\}_n$ using the proposed method with $\alpha=0.01$ and $\gamma^2=100$. The solid blue (resp. dashed red) curves depict the estimated (resp. actual) ranges. The data used to generate this figure are the same as for Fig. \ref{fig:single_pix5}.}
	\label{fig:single_pix6}
\end{figure}

To highlight the benefits of our online approach over batch-based methods we also consider the single-pixel measurements used in Fig. \ref{fig:single_pix5} and compare our approach to the classical cross-correlation method (see details in \cite{altmann2016lidar}). This approach is chosen as it is the fastest batch-based method which processes all the pixels independently. Although the comparison could have been performed using image sequences and more advanced methods, the competing methods would have led to significantly higher computational costs. To apply the cross-correlation, we first discretise the detection events uniformly over $\left[0;T_r\right)$ with a stepsize of $1$, which is much smaller and $s^2=200$ such that the discretisation bias can be neglected. For each batch of $N_0$ frames, the depth is then estimated by finding the delay that maximises the cross-correlation between the histogram of times of arrival within this batch and the discretised version of $s(\cdot)$. Fig. \ref{fig:single_pix6} compares the depth estimates obtained via cross-correlation for batches of $N_0=10$, $N_0=50$ and $N_0=100$ frames to those obtained using O3DSP. While small values of $N_0$ can lead to more accurate instantaneous estimates of the ranges, this figure shows that the results are also more sensitive to background detections due to the small number of detections within each batch of $N_0$ frames. Note that in extreme cases where $\pi_{p,n}$ is small, there might even be no detection in some batches. Note also in the top plot of Fig. \ref{fig:single_pix6} that the performance of the cross-correlation method is affected by the relative amount of background detections (larger errors for $w_{p,n}=0.3$ than for $w_{p,n}=0.8$, i.e., for $n \in \left[600;1100\right]$). Here, we initialised the proposed method using weakly informative parameters but it could be initialised using a batch-based method, such as cross-correlation, with the first few frames to improve the convergence speed. 

\subsection{Analysis of static and dynamic 3D scenes}
\label{sec:subsec:videos}
\begin{figure}[ht!]
	\includegraphics[width=\columnwidth]{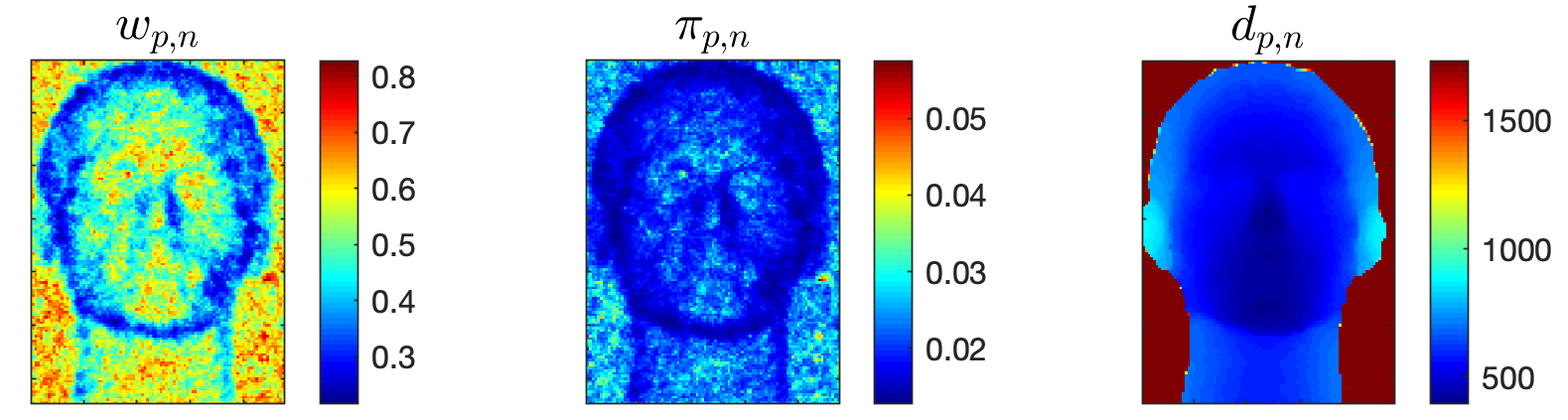}\\
		\vspace{-0.3cm}
	\caption{Ground truth parameters used for assessing the performance of the proposed method for reconstruction of a static scene.}
	\label{fig:GT_head}
\end{figure}
In this section, we first analyse the performance and convergence speed of O3DSP using simulated data based on real Lidar measurements conducted in \cite{altmann2016lidar,altmann2016target}. More precisely, we consider a series of $N=5000$ frames composed of $129 \times 95$ pixels and associated with a static scene whose range profile, probabilities of signal detection $\{w_{p,n}\}_p$ and probabilities of detection $\{\pi_{p,n}\}_p$ are depicted in Fig. \ref{fig:GT_head}. Here, we used $T_r=2500$. For most pixels, we have $\pi_{p,n}<5\%$, which corresponds to realistic observation conditions in the photon-starved regime. 

\begin{figure}[ht!]
	\includegraphics[width=\columnwidth]{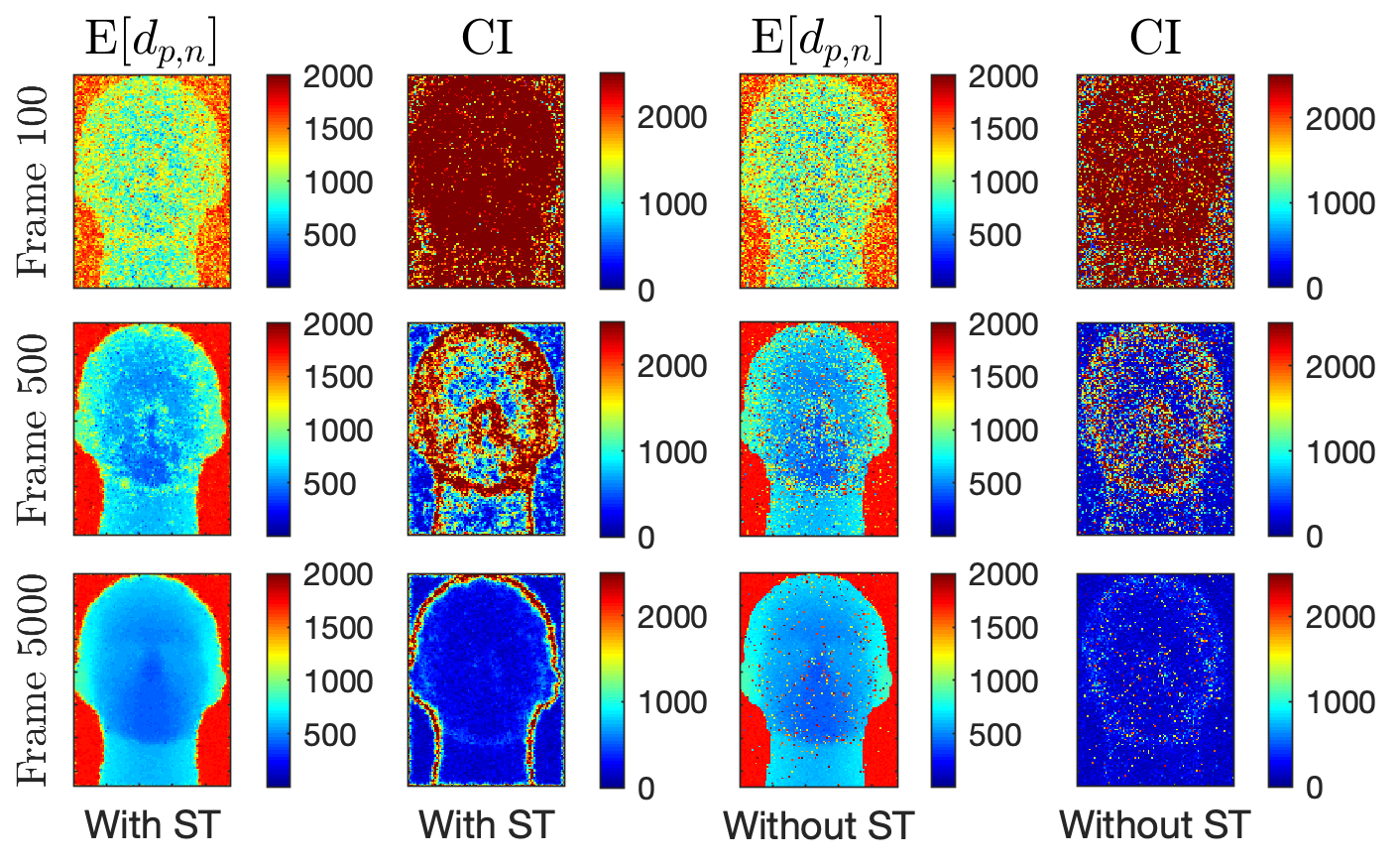}\\
		\vspace{-0.3cm}
	\caption{Online range estimation performance for a static scene: Estimated instantaneous means (first and third columns) and $\pm3$ standard deviation confidence intervals (CI) (second and fourth column) after $N=100$ frames (top), $N=500$ frames (middle) and $N=5000$ frames (bottom). The two columns on the left-hand side (resp. right-hand side) have been obtained with (resp. without) the proposed ST model.}
	\label{fig:head_depth}
\end{figure}

First, we compare the performance of O3DSP processing all the pixels independently, i.e., without smoothing of $\bar{\bfw}$ and with $\nu=1$ to the version using the proposed ST model. In this case, we used  $M=5$ neighbours, $\nu=0.99$ and $\bar{\bfw}$ was smoothed using a Gaussian filter with standard deviation $0.5$. In the two scenarios, we used $(\alpha,\gamma^2)=(0.1,10)$. Fig. \ref{fig:head_depth} depicts the estimated means and variances of the range estimates after 100, 500 and 5000 frames (top to bottom), with (left columns) and without (right columns) the ST model. These results illustrate the benefits of the ST model which improves the convergence speed of the algorithm and which reduces the number of isolated pixels with poorly estimated range (see bottom row of Fig. \ref{fig:head_depth}). O3DSP with the ST model is able to clearly identify regions of high uncertainty, i.e., the boundaries of the head where the range is likely to change suddenly, should the head move. Moreover, the uncertainty increases with the range difference between close pixels. For instance, the uncertainty is larger at the boundary of the head than in the neck/chin boundary.  

\begin{figure}[ht!]
	\includegraphics[width=\columnwidth]{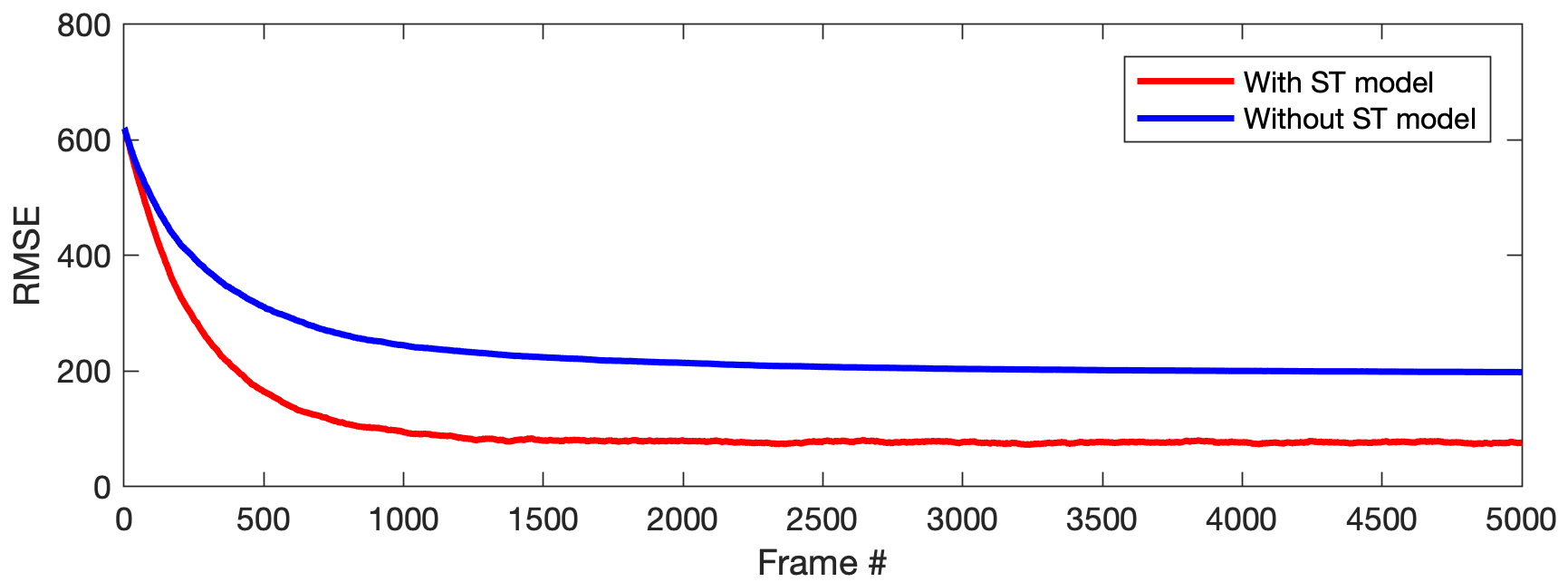}\\
		\vspace{-0.3cm}
	\caption{Range RMSEs obtained with (red lines) and without (blue lines) the proposed spatiotemporal (ST) model for the static scene considered in Fig. \ref{fig:GT_head}.}
	\label{fig:RMSE1}
\end{figure}

To assess quantitatively the convergence of the method, we use the range root mean square error (RMSE) defined as 
\begin{eqnarray}
\textrm{RMSE}_{n}=\sqrt{\dfrac{1}{P}||\hat{\bfd}_{n} - \bfd_n ||_2^2}, 
\end{eqnarray}
where $\bfd_n$ and $\hat{\bfd}_{n}$ are the actual and estimated range profiles in the frame $n$, respectively. Fig. \ref{fig:RMSE1} confirms that the proposed ST model improves the convergence speed and estimation performance in terms of RMSE. To ease the visualisation of these results, the generated data associated with this static scene, as well as the estimated range profiles are provided in a supplementary video associated with this paper (see Video 1 also available at \href{https://youtu.be/yxcJypmc_K4}{\nolinkurl{https://youtu.be/yxcJypmc_K4}}).

\begin{figure}[ht!]
	\includegraphics[width=\columnwidth]{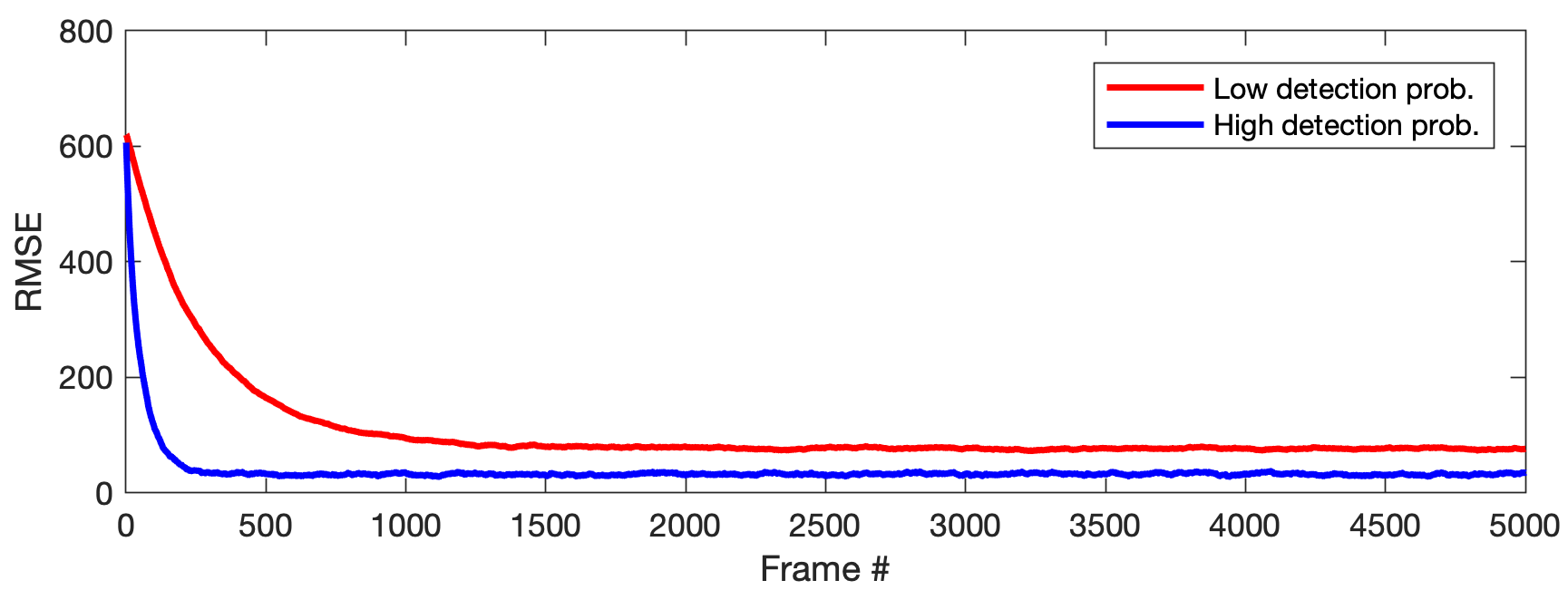}\\
		\vspace{-0.3cm}
	\caption{Range RMSEs obtained for the static scene using the parameters defined in Fig. \ref{fig:GT_head} (red lines) and by multiplying the probabilities of detection in Fig. \ref{fig:GT_head} (middle subplot) by 10 (blue lines).}
	\label{fig:RMSE_Pd}
\end{figure}

For completeness, we also generated data with the same parameters as above but with probabilities of detection $\{\pi_{p,n}\}_{p,n}$ multiplied by 10, when compared to those depicted in  Fig. \ref{fig:GT_head} (middle subplot), leading to an average probability of detection of $20\%$ per pixel and per frame. Fig. \ref{fig:RMSE_Pd} compares the convergence of the RMSEs for the original data (referred to as "low detection probability") and the new data set (referred to as "high detection probability"). As expected, increasing $\pi_{p,n}$ yields faster convergence and lower RMSEs at convergence due to the additional amount of (more frequent) detections available. 

Although O3DSP is designed to process sequences of individual detection events, these results illustrate also that it can be used to process ToA histograms associated with static scenes. Indeed, given the number or frames/laser repetitions used to build standard ToA histograms, it is straightforward to simulate $N$ frames of individual detection events. This particular case deserves a thorough comparison with existing batch-based methods, which is out of scope of this paper focusing of individual detection events. This should be considered in future work.
 
\begin{figure}[ht!]
	\includegraphics[width=\columnwidth]{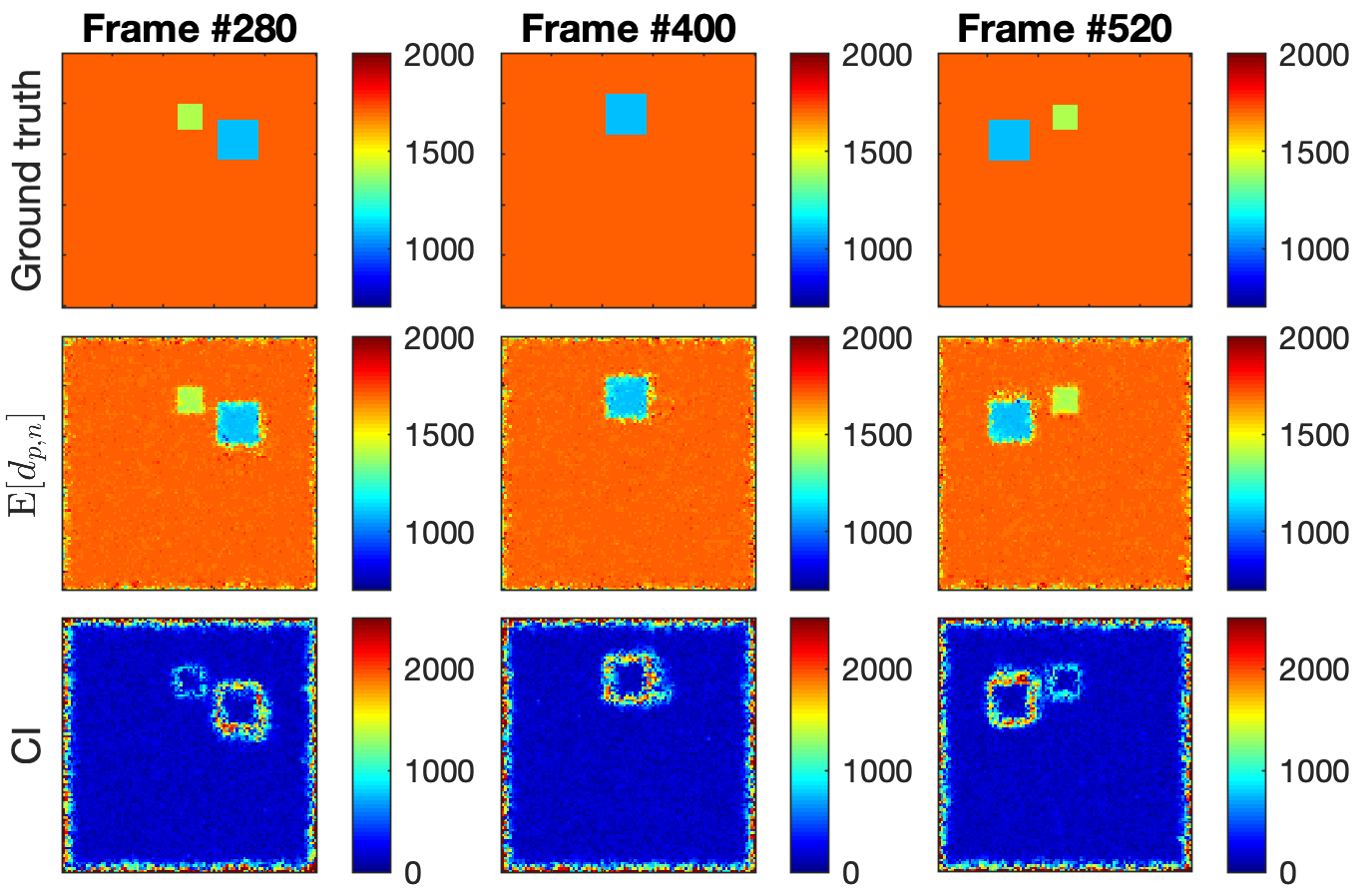}\\
		\vspace{-0.3cm}
	\caption{Example of range estimation for a dynamic scene with occlusion of one object. The full estimated range sequence can be seen in the supplementary Video 2.}
	\label{fig:disk}
\end{figure}

Finally, we applied our algorithm to the 3D reconstruction of a synthetically generated dynamic scene which consists of flat homogeneous rectangles, in front of a static backplane. For this experiment, we used $N=2400$ frames of $100 \times 100$ pixels with $\pi_{p,n}=0.5, \forall (p,n)$ and $T_r=2500$. During the first 800 frames, two objects are present. The first object is static while the second object describes a counterclockwise circular trajectory, centred at the centre of the image (rotation of $0.45^{\circ}$ per frame). During this rotation, the second object completely occludes the first one which then reappears. During the next 800 frames, the first object disappears suddenly and the second one describes the same trajectory as before (while its range remains unchanged) but its size varies. At frame 1600, a third object enters the field of view from the left and describes a horizontal movement (constant range), while the first object moves away from the backplane. Moreover, we set $w_{p,n}=0.5$ for the pixels associated with the backplane and $w_{p,n}=0.7$ for those associated with the two objects. This scenario is chosen to assess the robustness of the algorithm to occlusions and appearance of new objects. The parameters of the algorithm have been set to $M=5$, $\alpha=0.1$, $\nu=0.5$ and $\gamma^2=100$. The observed data as well as the estimated range profiles are provided in the second supplementary video associated with this paper (see Video 2 also available at \href{https://youtu.be/NYeoZ99BdtM}{\nolinkurl{https://youtu.be/NYeoZ99BdtM}}) . As an example, Fig. \ref{fig:disk} depicts estimated range profiles and associated uncertainties for three frames, namely before, during, and after the occlusion of one of the objects. Here, the range uncertainty is measured using the width on the confidence intervals (CI) defined as $6$ times ($\pm 3$) the standard deviations of the approximating Gaussians. For the three frames, we observe, as expected, higher uncertainties at the boundaries of the small rectangles. Moreover, this figure illustrates that the proposed method is able to recover occluded objects when they become visible again.

As mentioned in Sections \ref{sec:method} and \ref{sec:online}, one important property of the method is that, for a given frame, all updates (expect the smoothing step in line 11 of Algo. \ref{algo:algo1}) can be performed in parallel, using only estimates from one previous frame. In this work, the method has been implemented using Matlab 2017b running on a MacBook Pro with 16GB of RAM and a 2.9 GHz Intel Core i7 processor, leading to a average processing time of $4$ms per frame (with $P=10^4$ pixels).
\section{Conclusion}
\label{sec:conclusion}
In this work, we presented a first 3D reconstruction algorithm using individual photon detection events for online analysis of dynamic scenes. Based on assumed density filtering, the proposed method is computationally efficient as the data are processed partly in a parallel fashion (pixels in a given frame) and sequentially (successive frames), with a fixed computational cost.  The results presented in this paper have illustrated the flexibility and ability of the method to be used for static and slowly moving scenes (compared to the frame rate). Whilst the code has not been fully optimised, preliminary results conducted with a tailored implementation using a Titan Xp GPU indicate significant computational improvement (well below $1$ms per frame), paving the way to new and efficient streaming and processing of data directly from actual SPAD detector arrays.
While the proposed method is able to track relatively slow changes of the 3D profile, ongoing work include the development of more sophisticated models, able the better predict the dynamics of the 3D profile and in particular, sudden changes associated with the appearance of objects or the occurrence of new objects. This problem is also related to the potential presence of an unknown number of objects per pixel, as in \cite{Tachella2019_manipop} for instance, which should be addressed in future work, in particular for fast object detection. Although the range estimation does not seem to be significantly affected by the quality of the estimation of the probability of signal detection in the scenarios investigated, it would be also interesting to investigate in future studies whether the proposed methodology can be made more robust to extreme ambient illuminations where $w_{p,n}\ll1$. 

\section*{Acknowledgment}
This work was supported by the Royal Academy of Engineering under the Research Fellowship scheme RF201617/16/31, the ERC advanced grant C-SENSE, project 694888, and by the Engineering and Physical Sciences Research Council (EPSRC)  (grant EP/S000631/1) and the MOD University Defence Research Collaboration (UDRC) in Signal Processing. We gratefully acknowledge the support of NVIDIA Corporation with the donation of the Titan Xp GPU used for this research. The authors would also like to thank Dr Aurora Maccarone and Julian Tachella (Heriot-Watt University) for interesting discussions during the preparation of this work.

\ifCLASSOPTIONcaptionsoff
  \newpage
\fi



%
\bibliography{mybibfile}
\bibliographystyle{IEEEtran}

\end{document}